\documentclass{pasj00}
\SetRunningHead{Shigeyama et al.}{Early evolution of relativistic  ejecta}
\usepackage[pdftex]{graphicx}
\usepackage{times}

\begin{document}

\title{Early evolution of spherical ejecta expanding into the circumstellar matter at ultra-relativistic speeds}
\author{Toshikazu Shigeyama,\altaffilmark{1} Akihiro Suzuki\altaffilmark{1,2}, Ko Nakamura\altaffilmark{3}}
\altaffiltext{1}{Research Center for the Early Universe, Graduate School of Science, University of Tokyo, Bunkyo-ku, Tokyo 113-0033}
\altaffiltext{2}{Department of Astronomy, Graduate School of Science, University of Tokyo, Bunkyo-ku, Tokyo 113-0033, Japan}
\altaffiltext{3}{National Astronomical Observatory of Japan, Mitaka, Tokyo 181-8588, Japan}


\maketitle

\begin{abstract}
We present  a new self-similar solution describing early evolution of an ultra-relativistic flow resulting from a collision of homologously expanding spherical ejecta with the circumstellar matter, in which a shock wave propagates in the circumstellar matter while  a weak discontinuity propagates in the ejecta at the sound speed.
\end{abstract}
\section{Introduction}
Ultra-relativistic flows play important roles in some astrophysical phenomena like $\gamma$-ray bursts where fire balls with the Lorentz factors exceeding 100 are thought to drive the phenomena. There are several self-similar solutions involving such ultra-relativistic flows. \citet{Blandford76} found self-similar solutions for flows resulting from point explosions in the relativistic limit. Adopting their prescription, \citet{Nakayama05} described the shock emergence from a plane-parallel atmosphere. Furthermore, \citet{Nakamura06} found relativistic version of the spherical flow relevant to young supernova remnants in which ejecta interact with the circumstellar matter (CSM) through two shock waves and the contact surface in between. Here we present a new self-similar solution describing flows associated with ejecta expanding into the CSM in somewhat different manner. A shock wave propagating in the CSM leads the jet and forms a dense shell detaching itself from the homologously expanding ejecta. The inner wave front is a rarefaction wave propagating into the ejecta. This phase follows the spreading phase where the dense shell expands almost freely \citep{Meszaros93} and precedes the phase described by \citet{Nakamura06}.

A  similar flow has been observed in 2-D axisymmetric simulations for jets driven by aspherical supernova explosions \citep{Nagakura11, Suzuki11} though detailed analyses of the structure of the jet in the radial direction is prohibited by coarse grids to cover the large dynamic range. It is expected from the same calculation that the propagation of the ultra-relativistic jet  in the vicinity of the axis of symmetry is well approximated by the corresponding flow assuming spherical symmetry because the fluxes of mass,  momentum, and energy as well as the pressure gradient in the lateral direction are too small to affect the structure and propagation of the jet. Therefore a spherically symmetric solution presented in this paper will be useful to investigate the radial structure of jets near the axis of symmetry.

In the next section, we discuss initial conditions to realize the flow described by the self-similar solution presented in this paper. In section \ref{sec:urf}, we present the basic equations governing the ultra-relativistic flows in spherical symmetry. In section \ref{sec:ejecta}, we describe the density and pressure distributions of the ejecta.  In section \ref{sec:rarefaction}, we formulate the flow propagating in the ejecta using a prescription similar to that presented in \citet{Nakamura06}. In section \ref{sec:shock}, we discuss the propagation of the shock in the CSM using the self-similar solution found by \citet{Blandford76}. The boundary conditions at the contact surface are given in section \ref{sec:boundary}. Section \ref{sec:results} presents the results. In section \ref{sec:CD}, we conclude the paper.

\section{Initial conditions}\label{sec:init}
 In the non-relativistic limit, a collision of supersonic ejecta with the CSM results in two shock waves with a contact surface in between if the relative velocity $v_{\rm e}$ satisfies the inequality  \citep{Landau87}, 
\begin{equation}
v_{\rm e}>\sqrt{\frac{(\hat{\gamma}+1)P_{\rm e}}{2\rho_{\rm CSM}}},
\end{equation}
where $P_{\rm e}$ denotes the pressure in the ejecta, $\rho_{\rm CSM}$ the proper mass density of the CSM, and $\hat{\gamma}$ the adiabatic index, which is assumed as $\hat{\gamma}=4/3$ in the rest of the paper. This inequality usually holds in supernovae. In reality, young supernova remnants like Cassiopeia A are known to have such a structure \citep{Gotthelf01}. On the other hand, in the ultra-relativistic limit, a similar argument yields a different condition for leading to two shock waves, that is, 
\begin{equation}\label{eqn:cond}
\gamma_{\rm e}\equiv\sqrt{\frac{1}{1-v_{\rm e}^2}}>\sqrt{\frac{3P_{\rm e}}{4\rho_{\rm CSM}}}.
\end{equation}
Here the velocity is  measured in units of the speed of light $c$. 

To validate this condition, we solve the shock tube problems with parameters satisfying  and not satisfying the above condition and present the results in Figure 1. In the left panel, two fluids with the density $\rho=10^3$ are in contact at the position $x=0$ at the time $t=0$. The fluid  in the region $x<0$ having the pressure  $P=1$ moves in the positive $x$ direction with the Lorentz factor $\gamma_{\rm e}=100$, while the fluid in the region $x\ge 0$ is at rest and has the pressure $P=10^7$.  Therefore these satisfy the above condition  (\ref{eqn:cond}) for two shock waves and the resultant flow in the figure at  $t=0.5$ actually shows two shock waves. Here the plotted pressure is scaled by a factor of $10^{-7}$ and the velocity is measured in units of the speed of light. If the pressure in the right fluid is enhanced by a factor of two, the initial configuration does not satisfy the condition for two shock waves and results in the flow pattern presented in the right panel at  $t=0.5$.  However, it should be noted here that the criterion (\ref{eqn:cond}) does not accurately define the boundary of these two flow patterns and should be regarded as an approximate guide line to guess the resulting  flow pattern from initial conditions. 
\begin{figure*}[htbp]
\begin{center}
\includegraphics[width=16cm, bb =170 100 646 300]{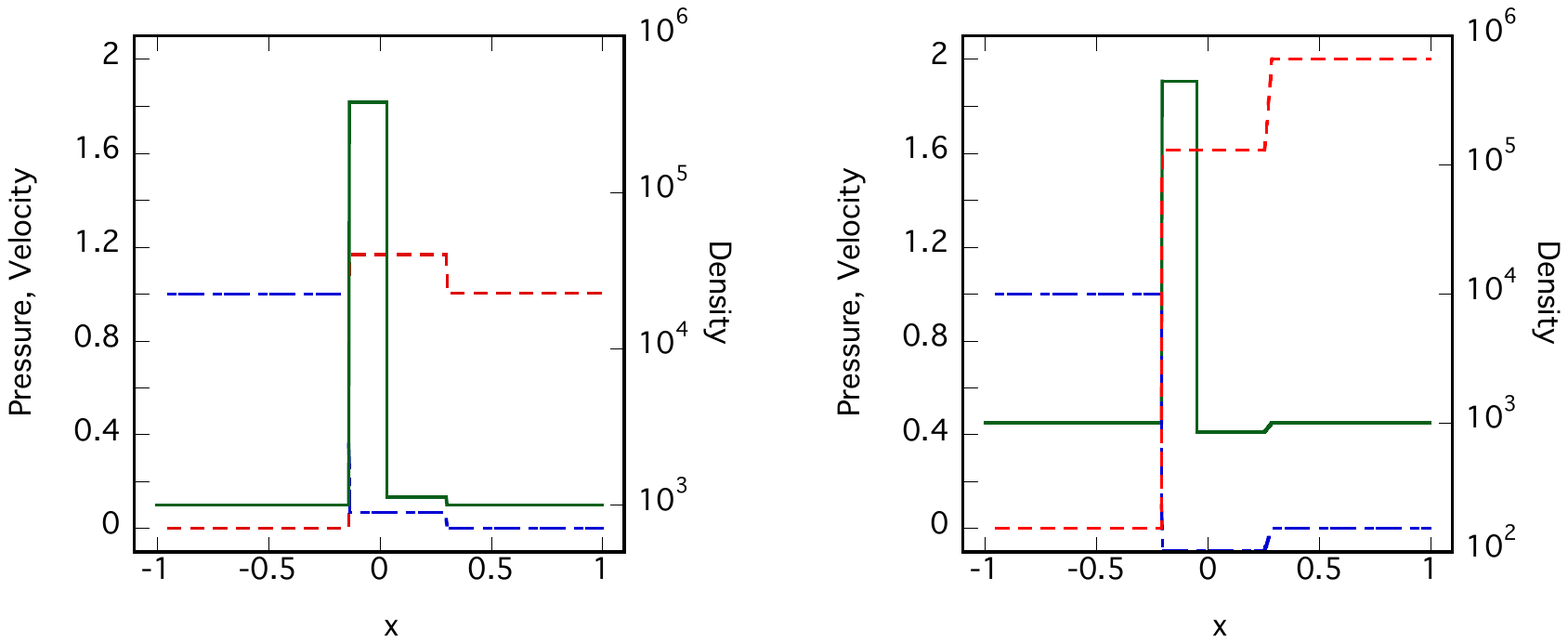}
\caption{Solutions for a relativistic shock tube problem with two different  sets of initial conditions. Distributions of pressure (dashed line), velocity (dash-dotted line), and density (solid line) at time $t=0.5$ are shown. The initial conditions at $t=0$ are as follows. The density is equal to $10^3$. The fluid in the region with $x<0$ moves with the Lorentz factor 100 in the positive $x$ direction and has the pressure equal to 1. The  pressure in $x\ge 0$ is $10^7$ in the left panel while $2\times10^7$ in the right panel. The pressure is scaled by a factor of $10^{-7}$. }
\label{shocktube}
\end{center}
\end{figure*}

The value of the right hand side of  this condition (\ref{eqn:cond}) could be estimated for a collision of ejecta with a stationary stellar wind in spherical geometry as
\begin{equation}\label{eqn:ncond}
\sqrt{\frac{3P_{\rm e}}{4\rho_{\rm CSM}}}\sim6\times10^3r_{13}T_7^2\sqrt{\frac{v_8}{\dot{M}_5}},
\end{equation}
where  $r_{13}$ denotes the distance from the explosion site in units of $10^{13}$ cm, $T_7$  the temperature of the ejecta in units of $10^7$ K, $v_8$ the wind velocity in units of $10^8$ cm s$^{-1}$ blown by the progenitor star, and $\dot{M}_5$ is the mass loss rate in units of $10^{-5}M_\odot$ yr$^{-1}$. If this condition is relaxed due to the decrease of the temperature when the forward shock propagates outward,  the reverse shock will be eventually generated. Therefore we will present a self-similar solution for the spherically symmetric flow composed of a shock wave propagating in the CSM and a rarefaction wave in the ejecta resulting from the collision. This phase precedes  the reverse shock phase described by the solution of \citet{Nakamura06}.   Since the scale of the structure is very small, i.e., on the order of $r/\Gamma^2$ where $\Gamma$ denotes the Lorentz factor of the shock wave, this solution describes fine structures of axisymmetric jets in the radial direction near the symmetry axis that multi-dimensional numerical simulations hardly resolve. 

\section{Ultra-relativistic flow}\label{sec:urf}
The spherically symmetric flow moving at speed $v$ is governed by the following equations in the relativistic limit \citep{Blandford76},
\begin{eqnarray}
&\frac{dP\gamma^4}{dt}= \gamma^2\frac{\partial P}{\partial t}, & \label{rhydro1}\\
&\frac{d\ln\left(P^3\gamma^4\right)}{dt} = -\frac{4}{r^{2}}\frac{\partial r^2v}{\partial r}, & \\
&\frac{\partial\rho^\prime}{\partial t} + \frac{1}{r^2}\frac{\partial r^2\rho^\prime v}{\partial r} = 0. &\label{rhydro3}
\end{eqnarray}
Here  $r$ and  $t$ denote the radial and time coordinates. The Lagrange derivative is described by $d/dt$. The pressure is denoted by $P$ and the ultra-relativistic equation of state $\epsilon=3P/\rho>>1$ has been assumed. The mass density in the fixed frame is denoted by $\rho^\prime$, which is related to the mass density $\rho$ in the co-moving frame as $\rho^\prime=\gamma\rho$, where $\gamma$ is the Lorentz factor of the fluid element. 

In constructing the self-similar solution, we follow the procedure of \citet{Nakamura06} with  different inner boundary conditions.

\section{Homologous  ejecta}\label{sec:ejecta}
Homologously expanding ejecta resulting from a shock emerging from the stellar surface have a power law structure in terms of the Lorentz factor of the fluid element as \citet{Nakayama05} and \citet{Kikuchi07} clearly showed.
The velocity $v_{\rm e}$ of homologous ejecta at  distance $r$ and time $t$ is defined  as,
\begin{equation}
v_{\rm e} = \frac{r}{t}.
\end{equation}
The corresponding Lorentz factor is
\begin{equation}
\gamma_{\rm e} = \frac{1}{\sqrt{1-v_{\rm e}^2}}.
\end{equation}
The mass density of the ejecta follows the power law in terms of the Lorentz factor as,
\begin{equation}
\rho_{\rm e}=\frac{a}{t^3\gamma_{\rm e}^n},
\end{equation}
where a non-dimensional parameter $n$ and a constant $a$ have been introduced.  The  pressure is assumed to have a polytropic expression with a varying coefficient $K(r)$,
\begin{equation}
P_{\rm e}=K(r)\rho_{\rm e}^{4/3}\left(r\right).
\end{equation}
In \citet{Nakamura06}, the pressure in the homologous ejecta was neglected because a strong shock wave propagates in the ejecta. In this paper, a rarefaction wave is assumed to propagate in the ejecta and we need to specify the pressure distribution. 

\section{Rarefaction wave}\label{sec:rarefaction}
Suppose that the front of the rarefaction wave is located at $r=R_1$ at time $t$. Then the boundary conditions are described as
\begin{eqnarray}
&\gamma_1^2=\gamma_{\rm e}^2\left(R_1\right), &\\
&\rho_1\gamma_1=\rho_{\rm e}\left(R_1\right)\gamma_{\rm e}\left(R_1\right), &\\
&P_1 = P_{\rm e}=K(R_1)\rho_{\rm e}^{4/3}\left(R_1\right).\label{p@e}&
\end{eqnarray}
The rarefaction front propagates at  a speed $V_{1}(t)$. Following the prescription of \citet{Blandford76}, the corresponding Lorentz factor $\Gamma_{1}$  is assumed to depend on time  as 
\begin{equation}
\Gamma_1^2=At^{-m},
\end{equation}
where $A$ and $m$ are constants. Therefore $R_1$ is approximated by
\begin{equation}
R_1(t) = \int_0^td\tau V_1(\tau)\sim t\left(1-\frac{1}{2(m+1)\Gamma_1^2}\right).
\end{equation}
Using this expression,  other variables in the ejecta at $r=R_1$ are expressed as
\begin{eqnarray}
&\gamma_{\rm e}^2(R_1)=(m+1)\Gamma_1^2, &\\
&\rho_{\rm e}(R_1) = \frac{a}{(m+1)^{n/2}t^3\Gamma_1^n}.&
\end{eqnarray}
The coefficient $K(R_1)$ is assumed to be $K(R_1)=K_0(t/t_0)^l$ where $K_0$, $t_0$, and $l$ are constants.
Using these dimensional variables at the front, non-dimensional variables $F(\xi),\,G(\xi),\,H(\xi)$ are introduced as follows.
\begin{eqnarray}
&P(r,\,t)=K(R_1)\rho_{\rm e}^{4/3}(R_1)F(\xi),& \\
&\gamma^2(r,\,t)=(m+1)\Gamma_1^2G(\xi),\label{g@e}&\\
&\rho(r,\,t)\gamma(r,\,t)=\rho_{\rm e}(R_1)\gamma_{\rm e}(R_1)H(\xi).&
\end{eqnarray}
Here a non-dimensional coordinate $\xi$ was defined as $\xi=[1+2(m+1)\Gamma_1^2](1-r/t)$ so that the rarefaction front is located at $\xi=1$.
Since the rarefaction wave propagates at the sound speed equal to $1/\sqrt{3}$ relative to the matter in the relativistic limit, the Lorentz factors of the rarefaction front and the ejecta there are related with each other as
\begin{equation}
\Gamma_1^2=\frac{3-\sqrt{3}}{2}\gamma_{\rm e}^2(R_1).
\end{equation}
Thus the parameter $m$ is evaluated as $m=1/\sqrt{3}$.
 
The conversion of coordinate system $(r,\,t)$ to $(\xi,\,\Gamma_1)$ is described as
\begin{eqnarray}
&t\frac{\partial}{\partial r} = -\left[1+2(m+1)\Gamma_1^2\right]\frac{\partial}{\partial\xi},&\\
&\frac{d}{d\ln t}=-m\frac{\partial}{\partial\ln\Gamma_1^2}+\left[\frac{1}{G}-(m+1)\xi\right]\frac{\partial}{\partial\xi}.&
\end{eqnarray}
If we retain only leading terms, the resultant hydrodynamical equations deduced from equations (\ref{rhydro1})-(\ref{rhydro3}) become three ordinary differential equations.
\begin{eqnarray}
&\frac{d\ln F}{d\xi} = \frac{12 [1-(m+1)\xi G]}{6[1+(m+1)\xi G]}\frac{d\ln G}{d\xi}& \nonumber \\
& +\frac{[m(4n-12)+6l-24]G}{6[1+(m+1)\xi G]},&\label{f@e}\\
&\frac{dG}{G^2d\xi} = \frac{
8(m+1)(m+2)\xi G+2m(4n-6)}{4\left\{[1+(m+1)\xi G]^2-3[1-(m+1)\xi G]^2\right\}}&\nonumber\\
&+\frac{12l-4m-32}{4\left\{[1+(m+1)\xi G]^2-3[1-(m+1)\xi G]^2\right\}},&\label{g@e}\\
&\frac{dH}{d\xi}=\frac{H\left[2\frac{d\ln G}{d\xi}-(mn-m-2)G\right]}{2[1-(m+1)\xi G]}.\label{h@e}&
\end{eqnarray}

\section{Flow in the shocked circumstellar matter}\label{sec:shock}
Finally, we will present equations governing the self-similar flow in the shocked CSM and the associated boundary conditions. These are identical to those in \citet{Blandford76}.
\subsection{Shock conditions in the CSM}
The mass density, Lorentz factor, and pressure change across a strong shock wave propagating at a speed $V$ by the following relations. 
\begin{eqnarray}
&\rho^\prime_2\equiv\rho_2\gamma_2 = 2\rho_{\rm i}\Gamma^2, &\\
&\gamma_2^2 = \frac{\Gamma^2}{2}, &\\
&P_2 = \frac{2\rho_{\rm i}\Gamma^2}{3},&
\end{eqnarray}
where the subscript 1 refers to the values in the shocked fluid at the shock front, the subscript i the values in the un-shocked external medium at the shock front.  $\Gamma\equiv 1/\sqrt{1-V^2}$ denotes the Lorentz factor of the shock front. To maintain the self-similarity, the Lorentz factor of the shock front $\Gamma$ needs to evolve as $\Gamma_{1}$ and is assumed to depend on time  as $\Gamma^2=Bt^{-m}$. Here $B$ is a constant. The mass density in the CSM is assumed to have a power law distribution as $\rho_\mathrm{i}=b r^{-k}$ where $b$ and $k$ are constants. Non-dimensional self-similar variables for the pressure $f(\chi)$, the Lorentz factor $g(\chi)$, and the mass density $h(\chi)$ are defined as
\begin{eqnarray}
&P=\frac{2\rho_{\rm i}\Gamma^2}{3}f(\chi), & \label{p@def}\\
&\gamma^2 = \frac{\Gamma^2}{2}g(\chi), & \label{g@def}\\
&\rho^\prime = 2\rho_{\rm i}\Gamma^2h(\chi), & \label{h@def}
\end{eqnarray}
where the similarity variable $\chi$ has been introduced as
\begin{equation}
\chi=\left\{1+2(m+1)\Gamma^2\right\}\left(1-\frac{r}{t}\right).
\end{equation}

\subsection{Self-similar flow}
By substituting the expressions (\ref{p@def})-(\ref{h@def}) into equations (\ref{rhydro1})-(\ref{rhydro3}), equations governing the self-similar flow in the shocked CSM are obtained as
\begin{eqnarray}
&\frac{d\ln f}{gd\chi}=\frac{4\{2(m-1)+k\}-(m+k-4)g\chi}{(m+1)(4-8g\chi+g^2\chi^2)}, & \\
&\frac{d\ln g}{gd\chi}=\frac{(7m+3k-4)-(m+2)g\chi}{(m+1)(4-8g\chi+g^2\chi^2)}, & \\
&\frac{d\ln h}{gd\chi}=\frac{2(9m+5k-8)-2(5m+4k-6)g\chi}{(m+1)(4-8g\chi+g^2\chi^2)(2-g\chi)}&\nonumber\\
&+\frac{(m+k-2)g^2\chi^2}{(m+1)(4-8g\chi+g^2\chi^2)(2-g\chi)}.& \label{h@a}
\end{eqnarray}
Because $m=1/\sqrt{3}$, the similarity solution is uniquely determined by specifying the parameter $k$.

\section{The flow at the contact surface}\label{sec:boundary}
At the contact surface, the density distribution has a discontinuity while the pressure and the velocity are continuously distributed. Since the pressures in equations (\ref{p@def}) and  (\ref{p@e}) need to evolve with the same power of time $t$, we obtain the relation
\begin{equation}
m(2n+3)+3(k+l)=12.
\end{equation}
Using this relation in equations (\ref{f@e}) and (\ref{g@e}), the dependences of these equations on the parameters $l$ and $n$ are eliminated.
The continuous distribution of pressure at the contact surface also relates the introduced constants as
\begin{equation}\label{eqn:k}
\frac{K_0a^{4/3}}{bBA^{2n/3}t_0^l}=\frac{2(m+1)^{2n/3}f(\chi_{\rm c})}{3F(\xi_{\rm c})}.
\end{equation}The denominators of (\ref{h@a}) and (\ref{h@e}) indicate
\begin{equation}\label{eqn:fg}
G(\xi_{\rm c})\xi_{\rm c} = \frac{1}{m+1}\quad {\rm and }\quad g(\chi_{\rm c})\chi_{\rm c} = 2,
\end{equation}
where $\xi_{\rm c}$ and $\chi_{\rm c}$ denote the coordinates of the contact surface.

From the continuity of the Lorentz factors at the contact surface, equations  (\ref{g@def}) and (\ref{g@e}) yield the ratio of Lorentz factors of the two wave fronts as
\begin{equation}
\frac{\Gamma}{\Gamma_1} = \sqrt{\frac{2(m+1)G(\xi_{\rm c})}{g(\chi_{\rm c})}}.
\end{equation}
The asymptotic behavior of the density near the contact surface is described as 
\begin{eqnarray}
&h\propto\left(\chi_{\rm c}-\chi\right)^\frac{\sqrt{3}k-1}{3\sqrt{3}k+1-12\sqrt{3}},& \\
&H\propto\left(\xi-\xi_{\rm c}\right)^{-2-\frac{4(n+1)}{3\sqrt{3}k+1-12\sqrt{3}}}.&
\end{eqnarray}
The former equation indicates that the density in the CSM diverges if $1/\sqrt{3}<k<4-\sqrt{3}/9$, otherwise the density in the CSM becomes zero at the contact surface. From the latter equation, when $k=0$, i.e., the CSM has a uniform density, the density in the ejecta diverges at the contact surface if $n<6\sqrt{3}-3/2$. When $k=2$, i.e., the CSM is formed by a stationary stellar wind, the density in the ejecta diverges at the contact surface if $n<3\sqrt{3}-3/2$ .

\section{Results}\label{sec:results}
\begin{figure}[htbp]
\begin{center}
\includegraphics[width=8cm, bb =150 50 646 600]{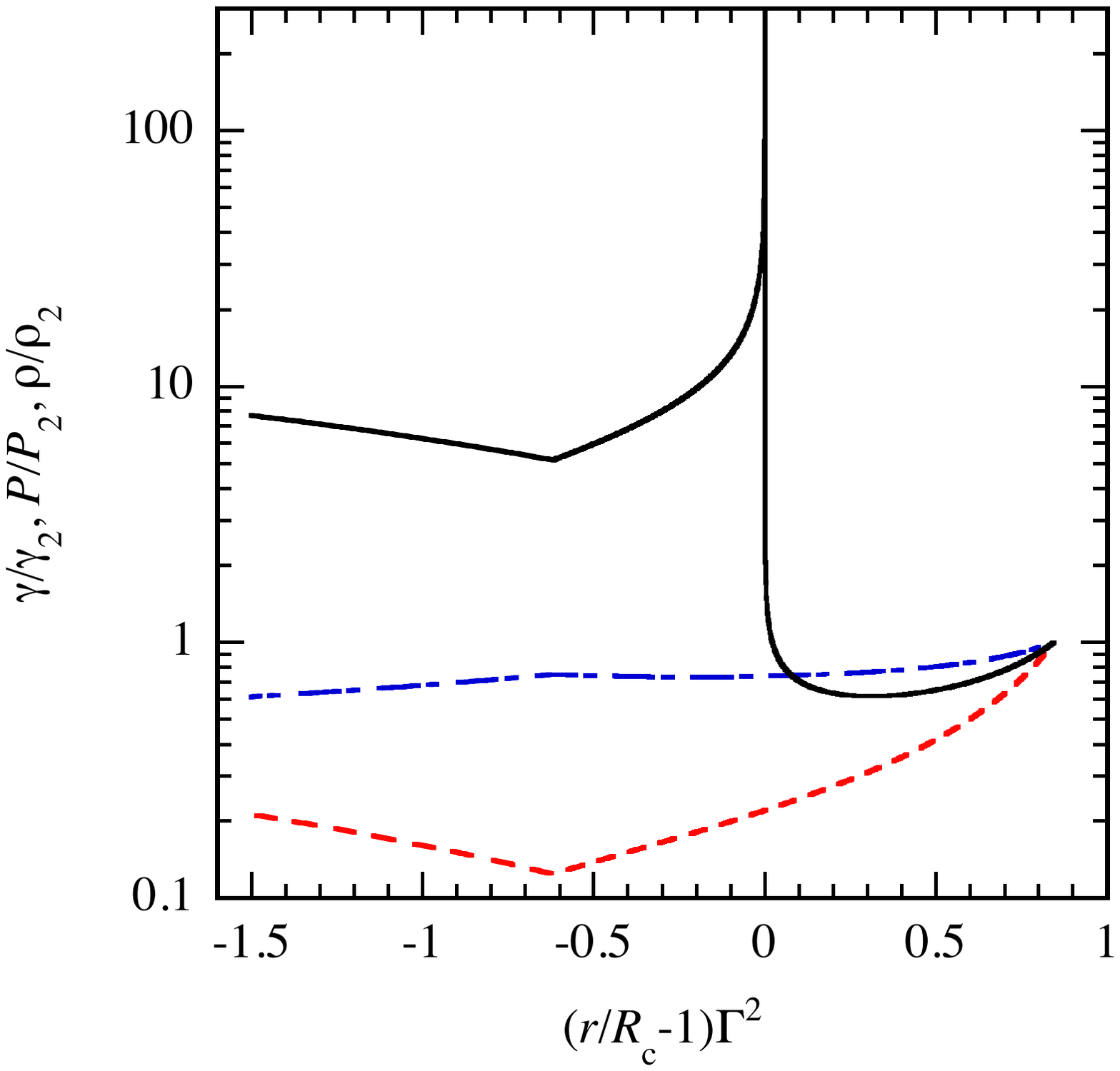}
\caption{Distributions of Lorentz factor (dash-dotted line), density (solid line), and pressure (dashed line) normalized with those values at the shock front as functions of scaled distance from the contact surface. The parameters are $n=2,\,k=2,\,l=2-7\sqrt{3}/9$.}\label{f2}
\label{profile}
\end{center}
\end{figure}

Results with a combination of parameters $n=k=2$ are presented in Figure \ref{f2}. 
The other parameters are as follows: $a=10^{6}$, $b=5.02\times10^{11}$,  $K_{0}=5.43\times10^{22}$, and $t_{0}=10^{4}$ in cgs units. These values are obtained from ejecta having the Lorentz factor of 300, the temperature of $10^{6}$ K, and the density of $10^{-10}$ g cm$^{-3}$ at the front. The ejecta collide at $r=ct_{0}$ with the stationary wind with the mass loss rate of $10^{-5}\,M_{\Sol}$ yr$^{-1}$. The wind velocity is assumed to be $1\times10^{8}$ cm s$^{-1}$. The plotted pressure, Lorentz factor, and density are normalized by those values at the shock front in the  CSM. Note that the density profile on the left side of the contact surface in this plot changes with time while the other profiles remain unchanged with time.

From the obtained solutions, we can derive a relation between the Lorentz factor and pressure of the ejecta and the density of the CSM near the wave as
\begin{equation}
\gamma_{\rm e}=\sqrt{\frac{3F(\xi_{c})g(\chi_{c})P_{\rm e}}{4f(\chi_{c})G(\xi_{c})\rho_{\rm CSM}}},
\end{equation}
by using equations (\ref{eqn:k}) and (\ref{eqn:fg}). The factor $F(\xi_{c})g(\chi_{c})/(f(\chi_{c})G(\xi_{c}))$ ranges from 2.2 to 4.5 for the parameter range of $0\leq k\leq 2$ with a fixed $n(=2)$, for example. Therefore the solutions satisfy the criterion (\ref{eqn:cond}) though we have rarefaction waves rather than reverse shock waves. Again, we should recognize the inequality (\ref{eqn:cond}) as an approximate guide line to classify the flow patterns resulting from  initial discontinuities partly because of the inhomogeneous structure of the flow.

\section{Conclusions and Discussion}\label{sec:CD}
We have presented a new self-similar solution describing the early phase of collision between spherical ejecta expanding at ultra-relativistic speeds and the CSM. The flow is characterized by the forward shock propagating in the CSM, the contact surface, and the rarefaction front with a weak discontinuity propagating in the ejecta.  As the forward shock proceeds in the CSM, a reverse shock eventually forms. Then another self-similar solution by \citet{Nakamura06} appropriately describes the flow afterwards.
 Though these solutions assume spherical symmetry, they can be used to describe flows of relativistic jets in the vicinity of the axis of symmetry since the 2-D numerical simulations show that the flux in the transverse direction is negligible there. 
 
When the rarefaction described by the present solution is transformed into the reverse shock depends on the structures of ejecta and the CSM.  If the ejecta are isothermal, the condition (\ref{eqn:cond}) together with equation (\ref{eqn:ncond}) suggests that the transition occurs at time $t\sim2\times10^{4}$ sec. This corresponds to $\sim$0.2 sec in the observer's frame. In reality, higher temperature or pressure in the inner ejecta will prolong the duration of this phase.  In the self-similar solution presented in this paper, the distribution of pressure $P(r,\,t)$ in the ejecta has a form of 
\begin{equation}
P\propto\gamma_{\rm e}^{2-4\sqrt{3}}t^{-4}.
\end{equation}
If the actual pressure distribution deviates from this and becomes shallower, then the transition is expected to occur.
The transition indicates a change of the time dependence of the shock Lorentz factor. It will be intriguing to investigate influences of this change on the emission in connection with light curve features of some gamma-ray bursts. For example, GRB 110205A exhibits a steep rise in its optical light curve starting at a few hundred seconds after the detection while the X-ray light curve shows a small bump around the same period \citep{Zheng11}. This might be related to the transition taking place at $r\sim 10^{18}$ cm from the explosion site and associated deceleration of the shock front, though calculations of the emission from such a shocked relativistic matter  is beyond the scope of the present paper.

\end{document}